\documentclass[twocolumn,aps,prb]{revtex4}
\usepackage{epsfig,graphicx}
\usepackage{amsmath,verbatim}
\usepackage{dcolumn, sidecap}

\begin{document}

\title{Pressure-Induced Phase Transformation in $\beta$-Eucryptite: an X-Ray Diffraction and Density Functional Theory Study}

\author{Yachao Chen,\textsuperscript{1}Sukriti Manna,\textsuperscript{2} Badri Narayanan\textsuperscript{1}\footnote{Present address: Nanoscience and Technology Division, Argonne National Laboratory, 9700 S. Cass Avenue, Argonne, IL 60439}, Zhongwu Wang\textsuperscript{3},
Ivar E. Reimanis,\textsuperscript{1}\footnote{Corresponding authors. Email: reimanis@mines.edu, cciobanu@mines.edu} and Cristian V. Ciobanu\textsuperscript{2$\dagger$}}

\affiliation{\textsuperscript{1}Department of Metallurgical and Materials Engineering, Colorado School of Mines, Golden, Colorado 80401, USA}

\affiliation{\textsuperscript{2}Department of Mechanical Engineering, Colorado School of Mines, Golden, Colorado 80401, USA}

\affiliation{\textsuperscript{3}Cornell High Energy Synchrotron Source, Cornell University, Ithaca, New York 14853, USA}

\begin{abstract}
Certain alumino-silicates display exotic properties enabled by their framework structure made of corner-sharing
tetrahedral rigid units. Using \textit{in situ} diamond-anvil cell x-ray diffraction (XRD), we study the pressure-induced transformation of $\beta$ eucryptite, a prototypical
alumino-silicate. $\beta$ eucryptite undergoes a phase transformation at moderate pressures, but the atomic structure
of the new phase has not yet been reported. Based on density functional theory stability studies and
Rietveld analysis of XRD patterns, we find that the pressure-stabilized phase belongs to the Pna2$_1$ space group.
Furthermore, we discover two other possible pressure-stabilized polymorphs, P1c1 and Pca2$_1$.
\end{abstract}

\maketitle

From their ubiquity in silica phases found in Earth's mantle\cite{mantle3, mantle2, mantle1} to their
importance for synthesizing ceramics with extraordinary electronic\cite{electronic} or mechanical properties,\cite{hard2, hard1} pressure-induced phase transformations in silicates provide fundamental insights into structure-property relations as well as ways to
control these relations so as to enable a variety of engineering applications.
Such applications range from transformation toughening in ceramics\cite{progtough}
to cathode materials for Li-ion batteries.\cite{LIB1, LIB2}
Amongst silicates, LiAlSiO$_4$ displays a wide variety of pressure-induced transformations.
Their structure consists of tetrahedra centered at Si or Al, connected in an anisotropic framework
that weakly binds lithium ions in the voids surrounded by the tetrahedra.\citep{Xu1999a}
This framework underlies a whole host of exotic phenomena and properties,
such as negative coefficient of thermal expansion,\cite{Tscherry1972a,Schulz1972a,Shin-ichi2004,Xu1999,Xu1999a,Morosin1975} negative compressibility, and one-dimensional ionic conduction.\citep{Schulz1972a,JohnsonJr1975,Alpen1977a,Press1980,Renker1983}
In addition, $\beta$-eucryptite (a LiAlSiO$_4$ polymorph) exhibits a pressure-induced phase transformation at pressures that are low enough to be exploited technologically.\cite{Jochum2009,Ramalingam2013}

The crystal structure of $\beta$-eucryptite (space group P6$_4$22 or P6$_2$22)
can be described as a stuffed derivative of $\beta$-quartz.\cite{Tscherry1972a,Schulz1972a,Shin-ichi2004,Xu1999}
For this structure, Morosin \textit{et al.}\citep{Morosin1975} reported a phase
transformation occurring at pressures as low as $\sim 0.8$GPa.
Zhang \textit{et al.} \citep{Zhang2002,Zhang2005} also found a new polymorph
(dubbed the $\epsilon$ phase) around 1~GPa, reporting that the transformation
from $\beta$ to $\epsilon$  was reversible; amorphization was found to occur at
pressures above 5~GPa.\cite{Zhang2002}
More recent indentation and Raman spectroscopy experiments revealed
that the critical pressure for the forward transformation ($\beta$
to $\epsilon$)  is higher than that for the reverse one,\citep{Jochum2009} suggesting
that it may be possible to produce metastable eucryptite phases under ambient conditions.

While there is no doubt that a phase transformation from
$\beta$-eucryptite to another crystalline phase exists at low pressures, the atomic structure
of the $\epsilon$ phase remains unknown. To date, the crystal system, lattice constants, and
Miller indices associated with the $\epsilon$ phase diffraction peaks reported by
Zhang \textit{et al.}\cite{Zhang2002} have not been reproduced by other groups.
Here, we have carried out x-ray diffraction (XRD) experiments and density functional
theory (DFT) calculations in order to elucidate the atomic structure of the $\epsilon$
phase. We have found that there are several polymorphs that become more stable
than the $\beta$ phase upon increasing pressure. From these,  we have determined that
the best match for the experimental XRD patterns is achieved by an
orthorhombic Pna2$_1$ structure.
Furthermore, we have found that the presence of Mg as a dopant leads to the observation of
two phases that coexist past the transition pressure, and identified their atomic structures
as well. This finding indicates a subtle competition between polymorphs under pressure,
and suggests a way to modify the number and type of coexisting phases
that could be used in transformation toughening.

In our experiments, $\beta$-eucryptite powders have been synthesized through a sol-gel route.\citep{Mazza1993,Mazza1994,Ramalingam2012a}
We have doped some of the samples with Mg, substituting for lithium up to 0.3mol\%.
XRD measurements under pressure have been performed \textit{in situ} at room temperature
in a diamond anvil cell (DAC), from ambient pressure to 5~GPa; more details are given in
Supplemental Material (SM).
The pressure $p$ is derived from changes in the ruby fluorescence\citep{Mao1986}
via $p$(GPa)=0.274$\Delta\Lambda$, where $\Delta\Lambda$(\AA) is the
difference between the ruby wavelength being detected and that at ambient pressure, 6942.1\AA.\citep{Yamaoka2012}
As evidenced by the lack of band splitting in ruby fluorescence, shear
was absent\cite{shear1, shear2}  in the sample, so all experiments have been carried out in hydrostatic conditions.

\begin{figure*}[htbp]
\begin{center}
\includegraphics[width=10cm]{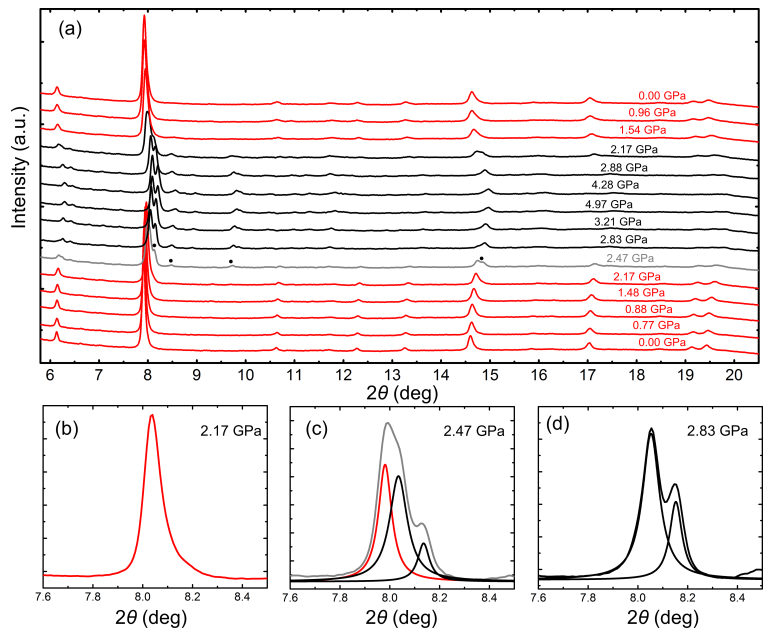}
\caption{XRD patterns showing the pressure-induced phase transformation in $\beta$-eucryptite.
(a) Variation of the pattern with pressure in a loading-unloading cycle.
The patterns that contain the new phase(s) are shown in black, while the $\beta$-phase is red.
(b)-(d) Detail views of the main peak at three pressures around the transition,
showing the main peaks (b) before, (c) during, and (d) after the transition.
}\label{fig:fig-1-XRDpatterns}
\end{center}
\end{figure*}

\begin{figure}[htbp]
\begin{center}
\includegraphics[width=7.5cm]{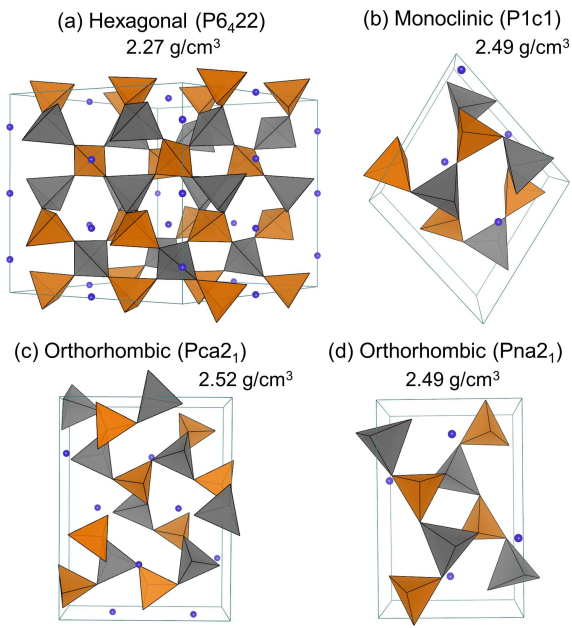}
\caption{Crystal structures of (a) $\beta$-eucryptite and (b)-(d) three denser polymorphs,
and their densities at zero pressure.
Al-(Si-)centered tetrahedra are shown in gray (tan), Li atoms are purple spheres.}\label{fig:fig-2-structures}
\end{center}
\end{figure}

\begin{figure}[htbp]
\begin{center}
\includegraphics[width=7.0cm]{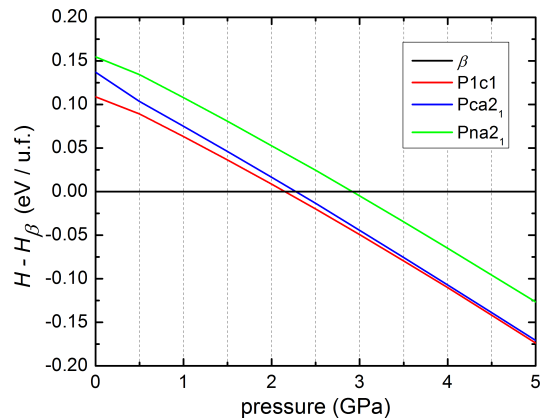}
\caption{Pressure-dependent enthalpy of LiAlSiO$_4$ polymorphs, with respect to the $\beta$-phase.
}\label{fig:fig-3-enthalpy}
\end{center}
\end{figure}

\begin{figure}[htbp]
\begin{center}
\includegraphics[width=7.5cm]{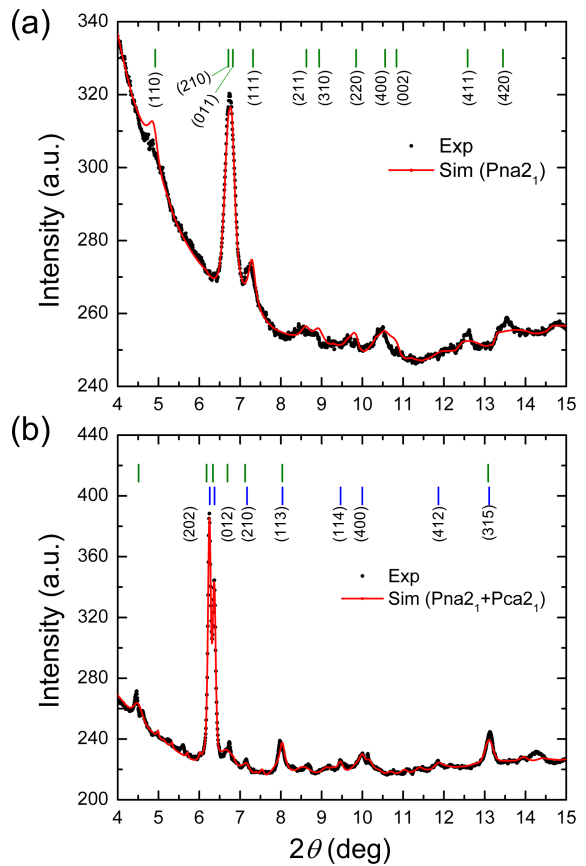}
\caption{(a) XRD pattern for the pure $\beta$-phase at 5~GPa, compared with the optimized simulated pattern for the Pna2$_1$ phase;
the other polymorphs have significant deviations from the experimental pattern.
(b) XRD pattern of the Mg-doped $\beta$-eucryptite at 5~GPa, compared with simulations of optimized mixture of Pna2$_1$ and Pca2$_1$.
For clarity, only the Miller indices corresponding to Pca2$_1$ peaks are shown in panel (b).
}\label{fig:fig-4-fitting}
\end{center}
\end{figure}

{\em In situ} DAC XRD experiments were conducted using an angle dispersive synchrotron
source at 
Cornell High Energy Synchrotron Source (CHESS).\citep{Wang2010}
Incident X-rays were converted by a double-bouncing monochromator
at a wavelength of $\lambda=$0.48596\AA.
Our data [Fig.~\ref{fig:fig-1-XRDpatterns}(a)] shows that at ambient pressure only the $\beta$ phase is present,
which has a dominant peak [Fig.~\ref{fig:fig-1-XRDpatterns}(b)]
corresponding to the (202) planes. Above a threshold pressure, this peak changes
shape as shown in Fig.~\ref{fig:fig-1-XRDpatterns}(c):  new, smaller peaks
appear at that pressure, and are marked by black dots on the 2.47GPa curve
in Fig.~\ref{fig:fig-1-XRDpatterns}(a). By fitting the complex peak shape at 2.47GPa
[Fig. ~\ref{fig:fig-1-XRDpatterns}(c)] with Lorentzian functions, we identify three
peak locations --one corresponding to $\beta$, and the other two corresponding to
the new phase(s). A small increase in pressure past 2.47 GPa leads to the
disappearance of the main $\beta$ peak, while the intensity of the other two peaks increases [Fig.~\ref{fig:fig-1-XRDpatterns}(d)].
Thus, for the Mg-doped samples analyzed, the transition occurs between 2.47GPa and 2.83GPa, with
the $\beta$ phase absent past 2.83GPa; pure samples exhibit somewhat smaller transition pressures.

We now focus on searching for crystalline phases that can become more stable
than $\beta$ upon pressure loading, and  hence could be the product
of the pressure-induced phase transformation of $\beta$.
To this end, we have analyzed over thirty structures made of corner-sharing AlO$_4$ and SiO$_4$
tetrahedra such that any Al-centered tetrahedron is surrounded by four Si-centered tetrahedra, and vice-versa.
These structures have been obtained in two ways: (a) by creating Li-stuffed derivatives of known SiO$_2$ structures,
similar to the way in which $\beta$-eucryptite is related to $\beta$-quartz; and (b) by converting the
alkaline atom M (M~=~Li, Na, K) of known MAlSiO$_4$ and MAlGeO$_4$ structures\cite{materialsproject} into Li,
and Ge into Si when necessary.
All structures have been relaxed using DFT in the generalized gradient approximation (GGA),\cite{DFTnote}
for each pressure from 0 to 5 GPa in increments of 0.5 GPa.
After eliminating duplicates, the crystals that we identified as having higher density than $\beta$ and low heat of formation
are shown in Fig.~\ref{fig:fig-2-structures}, with the space group and density at zero pressure
identified; $\beta$ is included as reference. The higher-density  phases also have lower symmetry, as
expected from the symmetry hierarchy of crystal systems.\cite{symhierarchy}
Two of the three new phases are orthorhombic, with space groups no. 33 (Pna2$_1$) and 29 (Pca2$_1$);
the remaining phase is monoclinic (group no. 7, P1c1).

Next, we discuss the relative stability of these phases with respect to $\beta$. In Fig.~\ref{fig:fig-3-enthalpy}, we
plot the difference between the enthalpy of each candidate phase and that of $\beta$, per unit formula.
Interestingly, past a certain  pressure each of these phases becomes
more stable than $\beta$ eucryptite (Fig.~\ref{fig:fig-3-enthalpy}); for these pressure-stabilized phases, the
lattice parameters are given in Table I, and the atomic  positions are given in SM.
Furthermore, all threshold pressures fall between 2~GPa and 3~GPa (Fig.~\ref{fig:fig-3-enthalpy}), in good agreement
with the experimental assessment of the range of pressures required for transition (Fig.~\ref{fig:fig-1-XRDpatterns}).
We cannot expect a more quantitative agreement because of the
uncertainties related to the indirect pressure measurements in the DAC, and
of the approximations made in the DFT calculations (for example, the
enthalpies are computed at 0~K and not at room temperature).
In order to identify more precisely the phases to which
$\beta$ transitions upon loading, we examine the XRD patterns at high-pressures.

We have performed Rietveld analysis of the XRD data using MAUD software,\cite{MAUD} designed
to refine the background, structural (atomic coordinates, occupancies, lattice parameters and angles),
and microstructural (particle size, lattice strain, residual stress, texture, etc) parameters via a least-squares method.\cite{MAUDpaper}
As input, we have supplied the XRD pattern and a candidate structure with its atomic coordinates and symmetry group.
The optimization proceeds with fitting the background and the peak shape; the shape of the peaks
was assumed to be of asymmetric, pseudo-Voigt form so it can fit crystal size  and strain broadening
of the experimental profiles.\cite{pV} Without refining the fractional atomic coordinates and thermal vibration parameters,
the following parameters have been optimized: background coefficients, scale, lattice
parameters, zero-shift error, and peak shape parameters.  The procedure yields a simulated optimized
spectrum for comparison with the experimental data.
We have carried out this procedure for each of the candidate structures [Fig.~\ref{fig:fig-2-structures}(b)-(d)].
The best fit for the XRD pattern of the pure, undoped sample is the Pna2$_1$ structure,
shown in Fig.~\ref{fig:fig-4-fitting}(a) for an XRD pattern  at $\sim 5$ GPa.

\begin{table}[h!]
\centering
 \begin{tabular}{l | c   c  c  c  c  c}
 \hline \hline
Phase   &$a$ (\AA) &$b$ (\AA) &$c$ (\AA) &$\alpha$ ($^{\rm o}$) &$\beta$ ($^{\rm o}$) & $\gamma$ ($^{\rm o}$) \\[0.5ex]
 \hline
$\beta$   & 10.575 & 10.575 & 11.391  & 90.000 & 90.000  & 120.000 \\
P1c1      & 8.200  & 8.288  & 5.147   & 90.000 & 90.000  & 106.290 \\
Pca2$_1$  & 10.079 & 5.033  & 13.114  & 90.000 & 90.000  & 90.000\\
Pna2$_1$  & 10.082 & 6.673  &  4.990  & 90.000 & 90.000  & 90.000 \\
 \hline
 \end{tabular}
\caption{Lattice parameters and angles for the crystal structures shown in Fig.~\ref{fig:fig-2-structures} computed
at zero pressure using GGA. The dependence on pressure is included in SM.}
\end{table}

It is worthwhile to compare this result with the previous reports on the phase transformations
of $\beta$ eucryptite. Morosin {\textit{et al.}} observed a phase transformation around 0.8~GPa, and
interpreted it as a hexagonal phase with lattice constants commensurate with
half of those of $\beta$.\cite{Morosin1975}
The structure of our monoclinic P1c1 phase is stabilized at the lowest pressure (Fig.~\ref{fig:fig-3-enthalpy}) and is
somewhat close to being hexagonal ($a$ and $c$ are within 1.1\% of each other, and the angle between them $\sim$12\% away from 120$^{\rm o}$),
so it is reasonable to infer that this phase may actually be the one found by Morosin {\textit{et al.}}\cite{MoroCant}
Zhang \textit{et al.} reported a reversible transformation around 1.5GPa to $\epsilon$ eucryptite:\cite{Zhang2002}
although these authors conclude that the $\epsilon$ phase is orthorhombic, its space group and the atomic structure have not been reported so far.
Our data also indicates that $\epsilon$ eucryptite is orthorhombic, but with different lattice constants; in addition,
we report here the space group (no. 33, Pna2$_1$) and the atomic structure [Fig.~\ref{fig:fig-2-structures}(e) and SM].
The previous indexing\cite{Zhang2002}  is not consistent with the reflection conditions
for most orthorhombic groups.\cite{bookSYMMETRY} On the other hand, the planes that we have found
in our XRD pattern [indexed in Fig.~\ref{fig:fig-4-fitting}(a)] obey the reflection conditions for group no. 33,
and are largely the same as those encountered in other materials with the space group Pna2$_1$.\cite{otherPna21a,otherPna21b}

We have also analysed XRD data of 0.3m\% Mg doped samples such as those in Fig.~\ref{fig:fig-1-XRDpatterns},
and have found that no single phase results in a satisfactory fit to the XRD data.
Therefore, we are led to assume that in the presence of Mg, there
can be two or more phases present at high pressures. We have analyzed
phase mixtures using MAUD for all possible combinations of the phases shown in Fig.~\ref{fig:fig-2-structures}(b)-(d).
The best fit for the XRD pattern of the Mg-doped eucryptite at 5GPa [shown in Fig.~\ref{fig:fig-4-fitting}(b)] is obtained from a
mixture of 45.3\% (by volume) Pna2$_1$ and 54.7\% Pca2$_1$.
It is rather intriguing that the presence of Mg in very small amounts
facilitates the occurrence of more than one crystalline phase upon pressure loading --which may be due to
entropic (mixing) or kinetic factors.

In summary, we have used XRD measurements and DFT calculations to elucidate the
structure of the pressure-stabilized $\epsilon$ phase of eucryptite.
While this answers a long-standing question, deeper investigations
are necessary to understand how to control the presence of different
polymorphs at low and moderate pressures.
The reversible nature $\beta\rightarrow \epsilon$ transition, coupled with
moderate values of the transition pressure (2-3 GPa) and of volume change
($\sim$9\%), make it suitable for transformation toughening of ceramic composites
for various applications. Knowledge of the $\epsilon$ phase
can enable a better control over the design of such composites
(\textit{e.g.}, optimizing the particle size, and distribution of
$\beta$-eucryptite particles to be employed in a suitable
matrix). Furthermore, we have found that the presence of small amounts of dopants
facilitates the coexistence of distinct polymorphs under pressure, which
may lead to novel properties displayed by such phase mixtures.

{\em Acknowledgments.} We gratefully acknowledge the support of U.S. Department of Energy's
Office of Basic Energy Sciences through Grant No. DE-FG02-07ER46397 and that of the National Science Foundation
for CHESS under Grant No. DMR-1332208.

{\small

\begin{thebibliography}{99}
\bibitem{DFTnote}The DFT calculations were performed with VASP\cite{VASP} at the level of
generalized gradient approximation using the Perdew-Burke-Ernzenhof exchange-correlation functional.\cite{PBE}
We have used projector-augmented pseudopotentials,\cite{Joubert} a 500 eV plane-wave energy cutoff, and
$3\times3\times3$ $k$-point grid for supercells with lattice parameters  $\geq 10$\AA\ (primitive cells
repeated accordingly). The atomic and lattice paramater relaxations were stopped when residual forces
became smaller than 0.01 eV/\AA.
\end{thebibliography}

}

\end{document}